\input harvmac
\lref\Mand{B.B. Mandelbrot, J. Business 36, 394-419 (1963).}
\lref\StanleyB{P. Gopikrishnan, V.Plerou, L.A.N.Amaral, M.Meyer, and H.E.Stanley, Phys. Rev. E 60, 5305-5316 (1999). }
\lref\StanleyC{P. Gopikrishnan, V.Plerou, L.A.N.Amaral, Xavier Gabaix, and H.E.Stanley, Phys. Rev. E 62, 3023-3026 (2000).}
\lref\StanleyD{P. Gopikrishnan, V.Plerou, Xavier Gabaix, and H.E.Stanley, Phys. Rev. E 62, 4493-4496 (2000). }
\lref\StanleyE{P. Gopikrishnan, V.Plerou, Xavier Gabaix, and H.E.Stanley, Phys. Rev. E 66, 027104 (2002). }
\lref\StanleyF{P. Gopikrishnan, V.Plerou, and H.E.Stanley, Nature 421, 130 (2003). }
\lref\Itzykson{C. Itzykson and J-M. Drouffe, Statistical field theory, Cambridge Univ. Press (1989).}
\lref\Watts{D.J. Watts and S.H. Strogatz, Nature 393, 440 (1998).}

\def\<{\langle}
\def\>{\rangle}

\Title{cond-mat/0402511}
{\vbox{\centerline{Critical Ising Model and Financial Market}}}
\smallskip
\centerline{Takeshi Inagaki}
\smallskip
\centerline{\it Yamato Laboratory, IBM Japan}
\smallskip
\centerline{\it Shimotsuruma, Yamato-shi, Kanagawa, Japan}

\bigskip
\bigskip

\noindent
We investigate Ising model description of dynamics of stock price. The model is defined in near 2 dimensions, one dimension is time and another represents ensemble of stocks, and strength of response of investors to price change corresponds to inverse temperature of the system. At critical temperature, infinitely long correlation among number of trades along time is observed and power-law tail in distribution of price fluctuation appears.
\vskip 3cm
\noindent
\Date{Feb, 2004}

\newsec{INTRODUCTION}
Emergence of power-law tail in variety of distribution function \refs{\Mand} in economic system gets much attention of physicists because it suggests existence of underlying critical phenomena. Especially, there exist amount of data that record stock trades over decades and fluctuation of stock price is one of most intensively studied topic. In this paper we study change of stock price
\eqn\Gdef{
G(t)=lnS(t+1)-lnS(t)
}
where $S$ is stock price. We try to explain observed power-law tail of the distribution function $P(G)$ with power-law exponent $-(\alpha+1)$ where $\alpha \approx 1.5$ for small $g$ ($0.5<g<3$) and $\alpha \approx 3$ for large $g$ ($3<g$) where $g$ is price change at individual trade \refs{\StanleyB}.  
\newsec{REVIEW OF FINDINGS IN EMPRICAL STUDIES}
Here is summary of finding in empirical studies \refs{\StanleyC, \StanleyD, \StanleyE, \StanleyF} of statistical properties of financial time series in stock market. Denote number of trades in a certain interval $\Delta t$ at time $t$ by $N_{\Delta t}(t)=N_0+\delta N_{\Delta t}(t)$ where $N_0$ is average value of $N_{\Delta t}$ over long term. Number of trades is strongly correlated along time as
\eqn\Ncorr{
\<\delta N_{\Delta t}(t+T) \delta N_{\Delta t}(t)\> \propto T^{-\gamma}
}
where observed value of $\gamma$ is around $\gamma \approx 0.3$. Another quantity measures activity in a market is share volume. Define amount of share volume of $n$ trades
\eqn\Qdef{
Q_{n}=\sum_{i=1}^{n} q_{i}
}
where $q_{i}$ is share volume at $i$th transaction. It was found that each $q_i$ is weakly correlated and distribution $P(Q_{n})$ of $Q_{n}$ is Levy stable distribution with a characteristic function
\eqn\Levy{
\phi(k)=e^{-m|k|^\alpha} 
}
where $\alpha\approx1.45$. Naively thinking, price changes of stocks are related to share volume $Q_{N_{\Delta t}}(t)$ share traded during $\Delta t$. More preciously, it is related to imbalance of share volume traded seller initiated and traded buyer initiated. That is defined by
\eqn\OmegaDef{
\Omega_{\Delta t}(t) = \sum_{i=1}^{N_{\Delta t}} a_i q_i
}
where $a_i=\pm 1$ is sign for seller initiated ($+1$) and buyer initiated ($-1$) trade. The function form of $\Omega \to G$ varies by the time scale of interval $\Delta t$. But the function form of $\Phi \to G$ is stable where
\eqn\PhiDef{
\Phi_{\Delta t}(t) = \sum_{i=1}^{N_{\Delta t}} a_i
}
(difference between number of buyer initiated and seller initiated trades in interval $[t, t+\Delta t]$) and it is approximated by
\eqn\Gfla{
\<G_{\Delta t}\>\propto tanh(A \Phi_{\Delta t})
}
where $A$ is a constant. Distribution function $P(\Omega)$ has a single peak around $\Omega=0$ if we pick trades with small local deviation of share volume
\eqn\SigmaDef{
\Sigma = \<|q_i a_i -\< q_i a_i \>|\>.
}
When $\Sigma$ exceeds a critical value $\Sigma_c$, peak of $P(\Omega)$ splits into two peaks with separation
\eqn\PsiDef{
\Omega=\pm|\Sigma-\Sigma_c|.
}
Because $q_i$ is $i.i.d$ process, it is expected that this behavior is inherited to $\delta N_{\Delta t}(t)$ as well.
\newsec{CORRELATION IN TIME}
Existence of correlation {\Ncorr} implies existence of interaction between $\delta N_{\Delta t}(t)$ and $\delta N_{\Delta t}(t+1)$. This interaction might be mediated by $\Phi_{\Delta t}(t) \to G_{\Delta t}(t)$. Because $q_i$ is Levy stable process {\Levy}, we get
\eqn\SigProp{
\Sigma \propto N_{\Delta t}(t)^{-{1\over\alpha}}
}
and
\eqn\PhiProp{
\Phi_{\Delta t}(t) \propto \pm |N_{\Delta t}(t)-N_c|.
}
From {\Gfla} and {\PhiProp},
\eqn\NtoGfla{\eqalign{
|\<G\>|&=zero \qquad \qquad \qquad \ \ (\delta N_{\Delta t}<<n_0), \cr
&\propto \delta N_{\Delta t}-n_0\qquad\qquad (n_0<\delta N_{\Delta t}<n_1),\cr
&= constant(\neq 0) \qquad \  (n_1<<\delta N_{\Delta t}).
}}
Except for range $n_0<\delta N_{\Delta t}<n_1$, $|\<G\>|$ behaves as a bi-level stepwise function of $\delta N_{\Delta t}$. If we assume causal relation $G_{\Delta t}(t) \to \delta N_{\Delta t}(t+1)$, probability distribution of $\delta N_{\Delta t}(t+1)$ is
\eqn\NtoNdetail{\eqalign{
P_{t+1}(\delta N_{\Delta t})&=p_0(\delta N_{\Delta t}) \qquad (\delta N_{\Delta t}(t)<n_0), \cr
&=p_1(\delta N_{\Delta t})  \qquad (n_1<\delta N_{\Delta t}(t)).
}}
\newsec{ISING MODEL}
For convenience, introduce a new variable
$$
x=ln(\delta N_{\Delta t})-ln({n_0+n_1\over 2})
$$
and assume symmetry between $p_0$ and $p_1$ of {\NtoNdetail} as
\eqn\psym{\eqalign{
p_0(-x)=p_1(x)\equiv p(x).
}}
Represent $\delta N_{\Delta t}$ by bi-level of states $|-\>$ for $x<0$ and $|+\>$ for $x>0$. Transition amplitude among states is
\eqn\NtoNdetail{\eqalign{
P_{++}&=P_{--}\equiv e^{+\beta}=\int_{0}^{\infty} dx \ p(x)\cr
P_{+-}&=P_{-+}\equiv e^{-\beta}=\int_{-\infty}^{0} dx \ p(x)
}}
where $\beta$ represents strength of correlation in time series. If $\beta=0$, there is no correlation among $\delta N_{\Delta t}(t)$. This leads to Ising like model with spin variable $s_i=|\pm\>$ on 1 dimensional lattice (time with a tick $\Delta t$) and Hamiltonian is given by
\eqn\Hamil{
H=\sum_{(i,j)} s_i s_j.
}
When $\beta < \beta_c$ the system is in the disordered phase and trades happen randomly. If $\beta > \beta_c$ it is in the ordered phase and spins are aligned on a direction. In this case mean value of $N_{\Delta t}$ is transiting and vacuum of the theory is unstable. If $\beta=\beta_c$ it is at a critical point and infinite length of correlation (power-law correlation) of spins (number of trade $\delta N_{\Delta t}$) appears. Correlation function of magnetization in Ising model is given by
\eqn\IsingCorr{
\<s_{t+\delta t} s_{t}\>=\delta t^{-(d-2+\lambda)}
}
where $\lambda={1\over4}$ \refs{\Itzykson }. Regarding observed correlation of $\delta N_{\Delta t}$, dimension of market Ising model is $d=2.05$. One dimension is time and other may emerge from the network structure of interaction among ensemble of stocks.   
Equation {\Gfla} suggests the analogy of the mean field
\eqn\GflaMag{
\<M\>\propto{e^{+M}-e^{-M}\over e^{+M}+e^{-M}}.
}
It seems interaction in the space dimension (stock ensemble) is not restricted to its nearest neighborhood, rather is mediated by the mean field.
\newsec{PRICE FLUCTUATION}
From correlation {\Ncorr}, variation of sum of $\delta N_{\Delta t}$ asymptotically behaves as
\eqn\Nasym{
\<\delta N_{\Delta t \times T}^2\>=\<\sum_{i=1}^{T} \delta N_{\Delta t}(i)^2\>\propto \alpha_0 T+\alpha_1 T^{2-\gamma}
}
where $\gamma \approx 0.3$ and $\alpha_0$ and $\alpha_1$ are constants of order $O(1)$. Around $T\approx 1$, this is approximated by
\eqn\DeltaNasym{
\<\delta N_{\Delta t \times T}^2\>\propto T^{1.35}.
}
From {\NtoGfla}, $\<G\>$ is propotinal to $\<\delta N_{\Delta t}^2\>^{1\over 2}$ for small $\<G\>$. For large $\<G\>$, size of individual fluctuation dose not depend on $\<\delta N_{\Delta t}^2\>^{1\over 2}$ and it behaves as the Gaussian due to the Central Limit Theorem. This results
\eqn\GflaAsym{\eqalign{
\<G^2\>^{1\over 2}&\approx \<\delta N^2\>^{1\over 2} \approx T^{1.35\over 2} \qquad \ (\<G\><1) \cr
&\approx \<\delta N^2\>^{1\over 4} \approx T^{1.35\over 4} \qquad \ (1<\<G\>).
}}
This diffusion behavior of $\<G^2\>^{1\over 2}$ is consequence of distribution of $G$ and we get observed power-law tail of the distribution function for $G$ as
\eqn\GflaPower{\eqalign{
P(G)&\approx G^{-({2\over1.35}+1)}\approx G^{-(1.48+1)} \qquad \ (\<G\><1) \cr
&\approx G^{-({4\over1.35}+1)}\approx G^{-(2.96+1)} \qquad \ (1<\<G\>).
}}
\newsec{CONCLUSION}
In this paper, we see possible application of the Ising model to modeling of causal relation of activity in financial market. The structure of interaction is still obscure. Small world networks \refs{\Watts} that interpolates between the Ising model and the mean field theory is one of candidates. 

\listrefs
\end